\documentstyle[11pt,paspconf,psfig]{article}

\begin{document}

\title{Superwinds and the Nature of Starburst Galaxies}
\author{M. D. Lehnert}
\affil{Max-Plank-Institut f\"ur extraterrestrische Physik, Postfach
1603, D-85740 Garching, Germany}

\begin{abstract}

I discuss the observational evidence that starburst galaxies are able
to drive galactic-scale outflows (``superwinds'') and then argue
generally that superwinds must have had an important role in galaxy
evolution.  To explore the role of feedback from massive stars, I
review results suggesting that starbursts seem to obey a limiting IR
surface brightness of about 10$^{11}$ L$_{\sun}$ kpc$^{-2}$,
corresponding to a maximum star-formation rate of about 45 M$_{\sun}$
yr$^{-1}$ kpc$^{-2}$ for a ``normal'' initial mass function.  I conclude
by discussing the role of winds in determining this upper-limit and
discuss recent results implying that winds might actually escape the
galactic potentials in which they reside.

\end{abstract}

\section{Introduction}

The processes that occur during the formation, evolution, and death of
massive stars are the engines which drive galaxy evolution.  While this
statement is obvious enough, this simple fact is often lost in the
quest to study the most spectacular objects or to discover the most
distant galaxy.  Because massive stars play such a fundamental role in
galaxy evolution implies that we must have an accurate and detailed
understanding of formation and evolution of massive stars and how the
processes of their interaction with their surroundings changes as they
evolve.

Massive stars are prodigious producers of ionizing photons and
mechanical energy and as such they regulate the ionization, physical,
and kinematic structure of ISM.  The generation of ionizing photons,
mechanical energy, and the cosmic rays by massive stars regulate the
ISM pressure, and by doing so, perhaps provide the mechanism by which
subsequent star-formation is ultimately regulated (``feedback'').
Specifically, it is this feedback from massive stars that ultimately
balances star-formation against gravitational instability, tidal sheer,
and dissipation that give galaxies the characteristics that they are
observed to have.  In all these areas, the details of stellar evolution
play a critical role in understanding how stars regulate all facets of
galaxy structure and evolution.  Hence determining parameters like mass
loss rates, time spent in various evolutionary stages, which stars go
supernova, all feed back into our understanding of the structure of the
ISM and hence how galaxies evolve.

Of course, perhaps the best way of directly observing the role of
massive stars in driving galaxy evolution is through the study of the
most intense star-forming galaxies in the universe -- starbursts.  The
purpose of this proceeding is to review the effects of high rates of
star-formation on the host galaxy's interstellar medium.  Such a
discussion demonstrates the central role that massive stars and the
effects of stellar evolution have on the properties of galaxies.  Of
course, demonstrating evolutionary effects directly is difficult.  We
will take an indirect course.  We first show that the preponderance of
observational evidence is in favor of starburst galaxies driving
galactic scale supernova (and stellar wind)-driven superwinds, that the
strength of the wind is dependent on the star-formation rate and
distribution, and then discuss the possible implications for
self-regulation of star-formation, how superwinds might drive galaxy
evolution, and review recent evidence that in fact, superwinds may escape
the galaxy potentials in which they reside.

\begin{figure}[!t]
\centerline{\psfig{figure=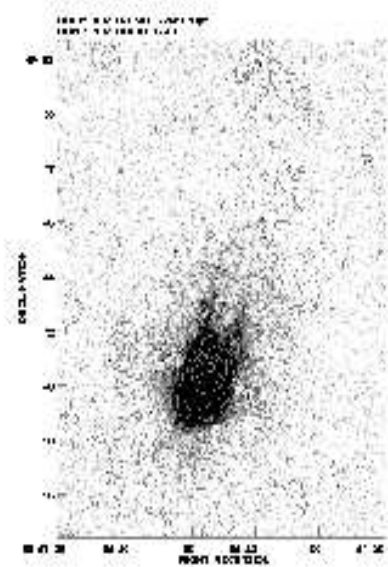,width=2.5in,height=3.5in}}
\vskip -0.5cm
\caption{The greyscale is the H$\alpha$ image and the contours are the
ROSAT PSPC image of M~82.  The spatial coincidence of the H$\alpha$ and
X-ray emission is quite good.  The faint ridge of emission to the north
is about 11.6 kpc (in projection) above the disk of M82 (see Devine and
Bally 1999).}
\end{figure}

\section{The Basic Physics of Winds}

Superwinds are thought to arise when the star-formation is intense
enough to create a region of rare gas of high temperature.  This hot
gas has an extremely  high pressure (T$\sim$10$^{7-8}$ K, n$_e$$\sim$
0.01 -- 0.001), much higher than the ambient ISM pressure and thus is
able to push the ambient ISM out preferentially in the direction of the
steepest pressure gradient (i.e., the minor axis in disk galaxies).
Given sufficient time and mechanical energy input, such a high pressure
region will eventually ``break out'' due to the various instabilities
(mainly dynamical ones) that cause the bubble walls to begin to
break-up.  When the bubble walls break apart, the wind begins to flow
outwards eventually reaching of-order its internal sound speed as a
free flowing wind (the effects of gravity can be safely ignored; see
Suchkov et al 1994; Tenerio-Tagle \& Mu\~noz-Tu\~n\'on 1997).  It may
also shock and accelerate ambient ISM clouds to velocities of several
hundred km s$^{-1}$ (Suchkov et al 1994).

\begin{figure}[!t]
\centerline{\psfig{figure=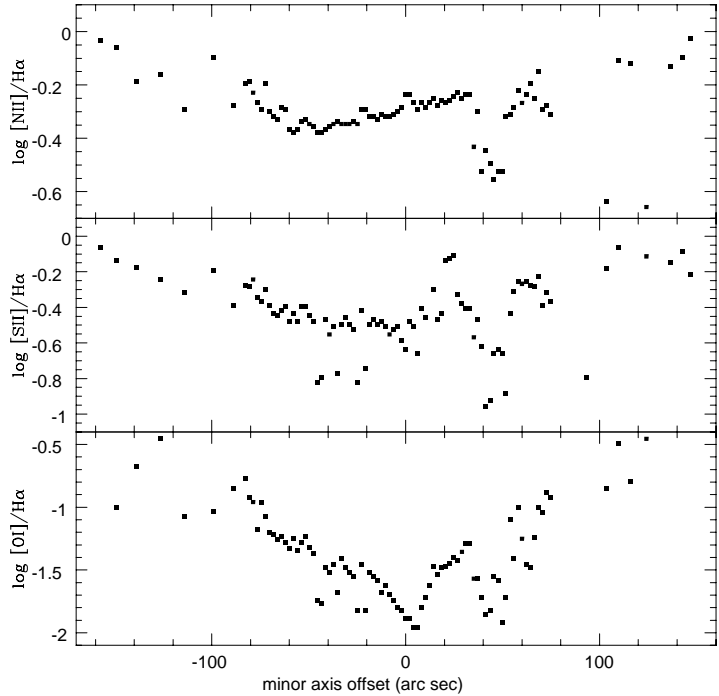,width=4.0in,height=4.0in}}
\caption{In this 3 paneled figure we show the line ratios as
a function of height above the disk of M82. As can be seen, the
line ionization line ratios increase with projected height above
the disk consistent with a greater contribution of shock-heating in
the gas far above the disk.}
\end{figure}

The conditions for establishing such a situation are not well
established theoretically and are only hinted at observationally.
Obviously a necessary (but not necessarily a sufficient condition) is
that the star-formation must be spatially concentrated.  This allows
the mechanical energy of the stellar winds and supernova remnants to be
effectively and efficiently thermalized.  If this mechanical energy is
not thermalized efficiently, or if lots of cool material is entrained
in the wind, it will radiate most of its energy away over a fairly
short time scale and thus will not sustain a flow for anything like a
sound crossing time over a galactic size scale (which is the minimum
condition necessary for driving a galactic scale wind).  Thus it is
apparent that there are many factors that can keep a galactic scale
wind from developing.

\section{Observational Properties of Winds}

Superwinds are a multi-wavelength phenomenon and the amount of
observational evidence at almost all wavelengths has been growing
tremendously over the past decade.  Different wavelengths probe
possibly different phases and phenomenology of the out-flowing wind but
all show ample evidence that starburst galaxies drive superwinds (e.g.,
Heckman, Armus, \& Miley 1990; Lehnert \& Heckman 1996a).  This evidence
includes: galactic scale bi-polar spatially extended soft X-ray
emission along the minor axis of starbursts disk galaxies (e.g., M82,
Fig.  1; Watson, Stanger, \& Griffiths 1984; Fabbiano 1988; Bregman,
Schulman, \& Tomisaka 1995; Moran \& Lehnert 1997; Ptak et al. 1997;
NGC~253, Fabbiano 1988; Persic et al. 1999; NGC~1569 Heckman et al.
1995; Della Ceca et al. 1996; NGC~2146 Armus et al.  1995; Della Ceca
et al. 1999; NGC~1808, Dahlem, Hartner, \& Junkes 1994; NGC~4449, Della
Ceca, Griffiths, \& Heckman 1997; NGC 3628, Dahlem et al. 1996; Arp 220,
Heckman et al. 1996; and for
a small sample of edge-on galaxies, Dahlem, Heckman, \& Weaver 1998),
extended galactic scale optical line emission with evidence for
shock-heating (Fig. 2; Lehnert \& Heckman 1996a; Heckman, Armus, \&
Miley 1990), emission line kinematics that often show split lines,
velocity offsets relative to systemic velocity, and broad lines in
the most extended emission line gas (Fig 3; Lehnert \& Heckman 1996a;
Heckman, Armus, \& Miley 1990), good correlation between ionization
state and line width in the extended gas (Lehnert \& Heckman 1996a),
nuclear optical emission line gas and X-ray emitting plasma with
extremely high pressures (several orders of magnitude higher than
ambient ISM pressure in the Milky Way) and with pressure profiles that
are consistent with that expected for out-flowing winds (Lehnert \&
Heckman 1996a;  Heckman, Armus, \& Miley 1990; Fabbiano 1988; e.g., all
of the X-ray references given previously), and extended polarized radio
emission (Dahlem et al.  1996).

\begin{figure}[!t]
\centerline{\psfig{figure=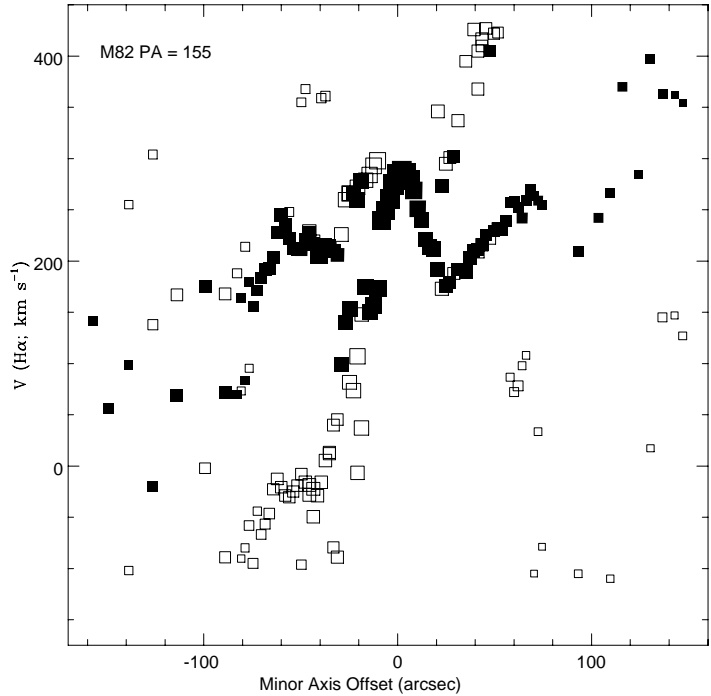,width=4.0in,height=4.0in}}
\caption{The kinematics of the extended H$\alpha$ emission along the
minor axis of M82
determined from a long slit spectrum from the KPNO 4m.
The black squares denote the highest surface gas.}
\end{figure}

Perhaps the best and most direct probes of the wind material are
X-rays.  As one might recall from the previous section, the wind fluid
should be hot in order to provide the high pressure necessary to drive
the outflow.  However, the predicted X-ray emissivity of the out-flowing
material is low (Suchkov et al. 1994; 1996) and yet the observed X-ray
luminosities of starburst galaxies are high (10$^{38}$ ergs s$^{-1}$
for a dwarf galaxy like NGC 1569; e.g., Heckman et al.  1995; Della
Ceca et al. 1996; to 10$^{42}$ ergs s$^{-1}$ for the almost
ultra-luminous IRAS galaxy NGC 3256; Moran, Lehnert, \& Helfand 1999).
A number of alternatives have been proposed to enhance the X-ray
luminosity produced by the wind. The wind could be centrally ``mass
loaded'' whereby quantities of ambient ISM could be mixed into the wind
in or near the starburst (Suchkov et al. 1996; Heckman et al. 1997).
Galactic halo clouds, whether pre-existing as might be there as tidal
debris of an interaction or tidal encounter or that have been carried
out from the disk with the wind material itself, could be evaporated or
ripped apart as they are overtaken by or interact with the wind fluid
(e.g., Suchkov et al. 1994).  Or the wind fluid could drive a shock
into a denser volume-filling galactic halo, with the observed X-ray
emission arising from the shocked-halo material rather than the
wind-fluid itself.  Mass loading the wind deep in the starburst nucleus
would likely produce a more uniform X-ray and optical line emission in
the halos of starbursts (dominated by adiabatic expansion and fluid
instabilities), while interactions with halo material would be expected
to produce very clumpy emission with large ranges of surface brightness
and temperatures.   Differentiating between the source of the halo
material is less straight-forward, perhaps by using HI as a tracer of
neutral material in the halo or perhaps determining the metal abundance
of the high surface brightness regions of X-ray emission.

\section{Conditions Necessary for Developing Outflows}

The theory of superwinds suggests that to generate a wind not only
requires active star-formation but in fact that the star-formation must
be intense.  Intense in this context means that the volume density of
energy input must be high enough so that the mechanical energy
deposited by stars into the ISM is effectively thermalized before it
has a chance to radiate away a significant fraction of its mechanical
energy.  Lehnert \& Heckman (1996) and Heckman, Armus, \& Miley (1990)
in studies of galaxies selected from the IRAS survey (i.e., infrared
bright) found that it is not only the star-formation rate but other
factors like the ``warmth'' of the IR emission and the ratio of IR to
optical luminosity also influenced the observational strength of the
wind.  Both of these additional factors are related to how enshrouded
the starburst region is and the UV heating rate of the dust (which are
proportional in some sense to the volume density of the energy input).
These results substantiate one of the basic tenets of the superwinds
hypothesis.

Specifically what these studies found was that to drive an outflow, a
galaxy should have large IR luminosities (L$_{IR}$ $>$ 10$^{44}$ erg
s$^{-1}$), large IR excesses (L$_{IR}$/L$_{OPT}$ $>$ 2), and warm
far-IR colors ($S_{60\mu m}/S_{100\mu m} \geq 0.5$).  Taking these
limits, the IR luminosity function, outflow rates $\propto$
L$_{IR}$ and using constants of proportionality determined in
well-studied examples like M82 and NGC253, the local space density of
galaxies, and a value of the Hubble time, we derive that superwinds have
carried out: M$_{eject}$ $\approx$ 5 $\times$ 10$^8$ M$_{\sun}$ in metals,
and E$_{KE \ and \ Thermal}$ $\approx$ 10$^{59}$ ergs per average
(Schecter L$^{\star}$) galaxy over the history of the universe.
Interestingly, these are approximately the mass of metals and the
binding energy of an average (L$^{\star}$) galaxy.  And this estimate
is conservative in that it assumes no evolution in the star-formation
rate with epoch.  Reasonable assumptions about the increasing interaction
rate and starburst number density with epoch would only increase our estimates
and demonstrates the potentially substantial role outflows have played in
galaxy formation and evolution.

\section{Questions Surrounding and Implications of Superwinds}

There are two central questions that studying superwind engenders.
Is there some critical star-formation intensity at which the mechanical
energy output from the massive stars halts further star-formation? 
Does the wind ultimately escape the potential of the
host galaxy?

Interestingly, it appears that the answer to the first question may be
yes.  For example, in a study of IR selected starbursts, Lehnert \&
Heckman (1996b) found that these galaxies have a ``large scale'' IR
surface brightness that appears to have a limit of $\approx$10$^{11}$
L$_{\sun}$ kpc$^{-2}$.  In a similar study but now including starbursts
with a range of selection methods and a wide range of redshifts, Meurer
et al. (1997) found a similar limit of L$_{BOL}$$\approx$ 2 $\times$
10$^{11}$ L$_{\sun}$ kpc$^{-2}$ (see Weedman et al. 1998 who suggest
that this limit may increase with increasing redshift -- about a factor
of 4 at redshifts $\approx$2--3).  This corresponds to a maximum
star-formation rate of about 45 M$_{\sun}$ yr$^{-1}$ kpc$^{-2}$ for a
``normal'' initial mass function.  This result should not be
over-interpreted.  It is not to mean that in all cases, star-formation
is stopped at this limit.  The cores of some HII regions and the
super-star clusters violate this ``limit''.  This is only to suggest
that perhaps the feedback from massive stars might provide a global
integrated limit to the star-formation rate.  On smaller scales,
different physical processes undoubtedly dominate compared to those on
larger scales which can lead to very high star-formation rates per unit
area or volume over limited scales.

In an analysis of this problem, Lehnert \& Heckman (1996b) argued that
this limit could plausibly be due to the out-flowing wind providing
enough pressure to overcome the hydrostatic pressure and thus halt
further star-formation.  To show this is plausible, they used M82 as a
test case and showed that star-formation in M82, which is at or near
this limit in the star-formation rate per unit area, does provide
enough pressure in the wind to balance the hydrostatic pressure of the
disk of M82.  However, such an hypothesis awaits further testing (see
also the analysis of Meurer et al. 1997).

With the rapid decline in the surface brightnesses of spatially
extended X-ray, optical, and radio emission, answering the question of
whether or not the wind fluid ultimately escapes the galactic potential
is currently a difficult question to answer definitively.  Certainly,
the model prediction is that it should, but whether or not it actually
does depends on where and how the wind is mass-loaded and how much cool
material ultimately mixes with the wind fluid.  However, some hints at
the answer are starting to emerge.  For example, Norman et al. (1996)
in a study of absorption lines in QSOs projected near the line of sight
to NGC 520 and NGC 253 found that it is plausible that the absorption
lines they detected along the line of sight to NGC 520 could be
associated with an outflow thus suggesting that the wind could reach
large distances from the galaxy.  However, as they pointed out, this is
not a unique interpretation of their results.

Another possible way of addressing this question has recently presented
itself.  Devine and Bally (1999) have recently discovered a ridge of
correlated X-ray and H$\alpha$ line emission 11.6 Kpc (in projection)
above the disk of M82 (Fig. 1).  In an analysis of this ridge of
emission, Lehnert, Heckman, \& Weaver (1999, submitted to ApJ) have
argued that this ridge represents the interaction of the superwind with
a dense cloud in the halo of M82 that may be part of the tidal debris
from the interaction between M82/NGC3077/M81 (Yun et al. 1993; 1994).
The analysis reveals that in order to explain the X-ray properties of
the ridge of X-ray/H$\alpha$ emission, the out-flowing wind must be
hitting the cloud at about 800 km s$^{-1}$ -- well in excess of the
escape speed at that distance above the plane for M82.  There is little
doubt that the wind in M82 is escaping its galactic potential.

\section{Conclusions}

We have demonstrated that the preponderance of evidence is strongly in
favor of starburst driving galactic scale superwinds.  In spite of the
fact that the existence of superwind is on a firm observational basis,
there are still many unanswered questions.  We have suggested that winds
are likely to have a huge impact during galaxy formation and subsequent
evolution but we do not understand the details of that statement.  We
know that the feedback from massive stars is likely to be important,
but how important?  What is the nature of the X-ray emission seen in
starburst galaxies?  The wind in M82 seems to be able to escape the
galactic potential, but is the escape of the wind plasma a general
feature of galaxies with superwinds?  Was it easier or more difficult
in the past for galaxies to drive winds and for these winds to escape
the galactic potential?  How will the details of stellar evolution and
the mechanical energy input change our views of winds and their
influence on galaxy evolution?  Some of these questions can only be
answered definitively when we have a complete understanding of the physics and
evolution massive stars including the role of Wolf-Rayet stars in
exciting and disturbing the ISM.

\acknowledgments
I would like to express my sincerest thanks to Karel van der Hucht for
his immense patience and understanding in waiting for my contribution.

\end{document}